\documentclass[twocolumn]{aastex62}

\received{-}
\revised{-}
\accepted{-}

\submitjournal{ApJ}

\shorttitle{ALMA and VLA Observations of EX Lupi}
\shortauthors{White et al.}

\begin{document}

\title{ALMA and VLA Observations of EX Lupi in its Quiescent State}

\correspondingauthor{J.A. White}
\email{jwhite@nrao.edu}

\author[0000-0001-8445-0444]{Jacob Aaron White}
\affiliation{National Radio Astronomy Observatory, 520 Edgemont Rd., Charlottesville, VA, 22903, USA}
\affiliation{Jansky Fellow of the National Radio Astronomy Observatory}
\affiliation{Konkoly Observatory, Research Centre for Astronomy and Earth Sciences, Konkoly-Thege Mikl\'os \'ut 15-17, 1121 Budapest, Hungary}
  
\author[0000-0001-7157-6275]{\'A. K\'osp\'al}
\affiliation{Konkoly Observatory, Research Centre for Astronomy and Earth Sciences, Konkoly-Thege Mikl\'os \'ut 15-17, 1121 Budapest, Hungary}
\affiliation{Max Planck Institute for Astronomy, K\"onigstuhl 17, D-69117 Heidelberg, Germany}
\affiliation{ELTE E\"otv\"os Lor\'and University, Institute of Physics, P\'azm\'any P\'eter s\'et\'any 1/A, 1117 Budapest, Hungary}  
  
\author[0000-0002-3446-0289]{A.G. Hughes}
\affiliation{Department of Physics and Astronomy,
University of British Columbia,
6224 Agricultural Rd.,
Vancouver, BC V6T 1T7, Canada}

\author[0000-0001-6015-646X]{P. \'Abrah\'am}
\affiliation{Konkoly Observatory, Research Centre for Astronomy and Earth Sciences, Konkoly-Thege Mikl\'os \'ut 15-17, 1121 Budapest, Hungary}
\affiliation{ELTE E\"otv\"os Lor\'and University, Institute of Physics, P\'azm\'any P\'eter s\'et\'any 1/A, 1117 Budapest, Hungary}

\author[0000-0002-4324-3809]{V. Akimkin}
\affiliation{Institute of Astronomy, Russian Academy of Sciences, Pyatnitskaya str. 48, 119017, Russia}

\author[0000-0003-4335-0900]{A. Banzatti}
\affiliation{Texas State University, Department of Physics, RFM Building 3227, 601 University Drive, San Marcos, TX 78666, USA}

\author{L. Chen}
\affiliation{Konkoly Observatory, Research Centre for Astronomy and Earth Sciences, Konkoly-Thege Mikl\'os \'ut 15-17, 1121 Budapest, Hungary}

\author[0000-0002-4283-2185]{F. Cruz-S\'aenz de Miera}
\affiliation{Konkoly Observatory, Research Centre for Astronomy and Earth Sciences, Konkoly-Thege Mikl\'os \'ut 15-17, 1121 Budapest, Hungary}

\author{A. Dutrey}
\affiliation{Laboratoire d'Astrophysique de Bordeaux, Universit\'e de Bordeaux, CNRS, B18N, All\'ee Geoffroy Saint-Hilaire, 33615, Pessac,France}

\author[0000-0002-9298-3029]{M. Flock} 
\affiliation{Max Planck Institute for Astronomy, K\"onigstuhl 17, D-69117 Heidelberg, Germany}

\author[0000-0003-3773-1870]{S. Guilloteau}
\affiliation{Laboratoire d'Astrophysique de Bordeaux, Universit\'e de Bordeaux, CNRS, B18N, All\'ee Geoffroy Saint-Hilaire, 33615, Pessac,France}

\author{A.S. Hales}
\affiliation{Joint ALMA Observatory, Avenida Alonso de C\'ordova 3107, Vitacura 7630355, Santiago, Chile}
\affiliation{National Radio Astronomy Observatory, 520 Edgemont Road, Charlottesville, VA 22903, USA}

\author[0000-0002-1493-300X]{T. Henning} 
\affiliation{Max Planck Institute for Astronomy, K\"onigstuhl 17, D-69117 Heidelberg, Germany}

\author{K. Kadam}
\affiliation{Konkoly Observatory, Research Centre for Astronomy and Earth Sciences, Konkoly-Thege Mikl\'os \'ut 15-17, 1121 Budapest, Hungary}

\author{D. Semenov}
\affiliation{Max Planck Institute for Astronomy, K\"onigstuhl 17, D-69117 Heidelberg, Germany}
\affiliation{Department of Chemistry, Ludwig Maximilian University, Butenandtstr. 5-13, 81377 Munich, Germany}

\author[0000-0002-8421-0851]{A. Sicilia-Aguilar}
\affiliation{SUPA, School of Science and Engineering, University of Dundee, Nethergate, DD1 4HN, Dundee, UK}

\author[0000-0003-1534-5186]{R. Teague}
\affiliation{Center for Astrophysics | Harvard \& Smithsonian, 60 Garden Street, Cambridge, MA 02138, USA}

\author[0000-0002-6045-0359]{E.I. Vorobyov}
\affiliation{University of Vienna, Department of Astrophysics, Vienna 1180, Austria}
\affiliation{Institute of Astronomy, Russian Academy of Sciences, Pyatnitskaya str. 48, 119017, Russia}

\begin{abstract}

Extreme outbursts in young stars may be a common stage of pre-main-sequence stellar evolution. These outbursts, caused by enhanced accretion and accompanied by increased luminosity, can also strongly impact the evolution of the circumstellar environment. We present ALMA and VLA observations of EX Lupi, a prototypical outburst system, at 100 GHz, 45 GHz, and 15 GHz. We use these data, along with archival ALMA 232 GHz data, to fit radiative transfer models to EX Lupi's circumstellar disk in its quiescent state following the extreme outburst in 2008. The best fit models show a compact  disk with a characteristic dust radius of 45 au and a total mass of 0.01 M$_{\odot}$. Our modeling suggests grain growth to sizes of at least 3 mm in the disk, possibly spurred by the recent outburst, and an ice line that has migrated inward to $0.2-0.3$ au post-outburst. At 15 GHz, we detected significant emission over the expected thermal disk emission which we attribute primarily to stellar (gyro)synchrotron and free-free disk emission. Altogether, these results highlight what may be a common impact of outbursts on the circumstellar dust.

\end{abstract}



\section{Introduction} \label{sec:intro}

Giant molecular clouds are the nurseries in which stars are born. The earliest phases of mass accumulation take place in the densest regions of the cloud cores and the rotation of these in-falling cores can lead to the formation of young stellar objects (YSOs) embedded in accretion disks.

A growing number of YSOs have been observed to have extreme outbursts \citep[e.g.,][]{audard14}, increasing the brightness by up to several magnitudes and the total luminosity by a factor of $10-100$. The prototypical outbursting YSO, FU Orionis, entered an outburst in 1936 \citep{herbig66} and is still slowly fading \citep{kenyon00}. This outburst has motivated the study of other similar systems (named FUors) that have also experienced outbursts of various timescales and intensities. In addition to the FUor class of young stars, there are systems that exhibit shorter duration and weaker outbursts called EXors (named after EX Lupi, discussed in detail below). Theoretical considerations \citep[e.g.,][]{zhu09, dangelo12, vorobyov15} suggest that EXors/FUors, and Sun-like stars in general, can build up a significant fraction of their mass during periods of enhanced accretion, referred to as episodic accretion. The outbursts of both EXors and FUors are thought to be due to this episodic accretion of material from their circumstellar disks onto the protostars. While it is generally accepted that the observed outbursts are due to accretion, the exact triggering and transport mechanism(s) that deliver material from the disk on to the star over short time periods typical for FUors/EXors is still unknown \citep{audard14}.  

Understanding the outburst mechanisms of protostars is important not only for building a complete picture of stellar evolution, but also for the potential implications for the planet formation process around low-mass stars.  If these types of outbursts are a common byproduct of the star formation process then they will undoubtedly also impact disk evolution and therefore the conditions in which planets form. Outbursts have been shown to potentially change the chemistry and mineralogy of the surrounding circumstellar disk \citep{abraham09, rab17, molyarova18}; could spur the growth of small solids through, e.g., evaporation and recondensation from a rapid evolution of the ice line \citep{cieza16}; and offer a potential solution to the luminosity problem \citep{hartmann96}.

EX Lupi is a young $1-3$ Myr M0 star (0.6 M$_{\odot}$, 0.7 L$_{\odot}$ quiescent luminosity) located $157.7\pm0.9$ pc away in the Lupus 3 cloud \citep{lombardi08, sipos09, frasca17, bailer18}. There have been multiple observed outbursts in the EX Lupi system, the most recent of which was in 2008. During this last outburst, EX Lupi was monitored in the optical and infrared. The outburst caused the annealing of amorphous silicate particles to crystalline grains, which were transported to the outer comet-forming zone \citep{abraham09, abraham19}. The direct effects of the outburst on the outer disk, however, are still yet to be fully characterized \citep[e.g.,][]{aguilar12}. The outer disk was only recently spatially resolved by \citet{hales18} using the Atacama Large Millimeter Array (ALMA) at 232 GHz. They found no indications of substructure in the disk and used radiative transfer modeling to constrain the total dust mass to $1.0\times10^{-4}~\rm M_{\odot}$. \citet{kospal20} used spectropolarimetric monitoring of EX Lupi post-outburst with the Canada-France Hawaii Telescope (CFHT) to place constraints on the stellar magnetic fields. Their preliminary analysis suggests the surface magnetic field is 3 kG, stronger than what has been observed for some embedded FUor/EXor objects \citep[e.g., 1 kG observed in FU Orionis][]{donati05} and more in line with what is seen in Classical T Tauri stars with field strengths of up to a few kG \citep{johnskrull07, bouvier07,johnskrull09}. The strength and stability of the stellar magnetic field in EX Lupi can influence the stability of the accretion columns \citep{aguilar15} and has significant implications on possible outburst mechanisms.

In this paper, we present long wavelength observations of EX Lupi. In Sec.\,\ref{sec:obs}, we describe the data and calibration procedures. In Sec.\,\ref{sec:models}, we outline the model fitting approach used and in Sec.\,\ref{sec:disc} we discuss the results. 

\section{Observations} \label{sec:obs}

\begin{table*}
\begin{center}
\caption{Summary of the observations. The ALMA 232 GHz data is from \citet{hales18} with no additional calibration procedures. The peak fluxes were measured from the CLEANed images and the total flux was calculated in the {\scriptsize CASA} task \textit{uvmodelfit}. The stated uncertainties do not include the absolute flux calibration uncertainty, which is $\leq10\%$ at these frequencies for ALMA and $\leq5\%$ for VLA. $^{*}$The 45 GHz VLA observations resulted in a non-detection, therefore we include a $3\sigma$ upper level limit on the peak flux.}\label{obs_overview}
\bigskip
\begin{tabular}{c | c c c c}
\hline

 & 232 GHz & 100 GHz & 45 GHz & 15 GHz \\
 \hline\hline
Facility & ALMA & ALMA & VLA & VLA \\
Observation Date & 2016 Jul 25 & 2018 Jan 27, Mar 17, Mar 19 & 2019 May 13, Jul 10 & 2019 Mar 09\\
Flux Calibrator & J1427-4206 & J1427-4206, J1517-2422 & 3C286 & 3C286 \\
Beam Size & $0\farcs32\times0\farcs26$ & $1\farcs09 \times 0\farcs81$ & $0\farcs36 \times 0\farcs16 $ & $1\farcs88 \times 0\farcs42$\\
Beam PA & $61.8^{\circ}$ & $87.8^{\circ}$ & $10.2 ^{\circ}$& $10.5^{\circ}$\\
$\sigma_{\rm rms}$ & 0.038 mJy beam$^{-1}$ & 0.013 mJy beam$^{-1}$& 0.043 mJy beam$^{-1}$ & 0.007 mJy beam$^{-1}$ \\
Peak Flux & 8.8 mJy beam$^{-1}$ & 1.98 mJy beam$^{-1}$ & $<0.130^{*}$ mJy beam$^{-1}$ & 0.044 mJy beam$^{-1}$ \\
Total Flux & $17.37\pm 0.15$ mJy & $2.72\pm0.013$ mJy & - & $0.050\pm0.008$ mJy \\

\end{tabular}
\end{center}
\end{table*}

The analysis presented here uses a combination of new radio observations with the \textit{NSF's Karl G. Jansky Very Large Array} (VLA), new ALMA 100 GHz observations, and ALMA 232 GHz data from literature \citep{hales18}. All of the observations are summarized in Table\,\ref{obs_overview} and the new ones are detailed below.  

\subsection{VLA Observations}

The VLA observations (ID 19A-145, PI White) were centered on EX Lupi using J2000 coordinates RA $ = 16^{\rm h}~ 03^{\rm m} ~  7.09^{\rm s}$ and $\delta = -40^{\circ} ~18' ~05\farcs10$. The Scheduling Blocks (SB) for each observation were executed on different days but all used the B configuration with 26 antennas and baselines ranging from 0.21 to 11.1 km.  Quasar J1607-3331 was used for bandpass and gain calibration. Quasar 3C286 was used as a flux calibration source. Data were reduced using the {\scriptsize CASA 5.4.1} pipeline, which included bandpass, flux, and phase calibrations \citep{casa_reference}. The absolute flux calibration of the VLA at these wavelengths is typically $\sim5\%$. \footnote{For a note on the VLA flux calibration uncertainty, see science.nrao.edu/facilities/vla/docs/manuals/oss/performance/fdscale. } All of the SBs used the same sources for calibration.

The 15 GHz data from the VLA were acquired in Semester 19A on 2019 March 09 and had a total on-source time of $948\rm~s$. The observations used a Ku Band tuning setup with $3\times2$ GHz basebands and rest frequency centers of 13 GHz, 15 GHz, and 17 GHz. This gives an effective frequency of 15 GHz (1.99 cm). The data were imaged with a natural weighting and cleaned using {\scriptsize CASA}'s \textit{CLEAN} algorithm down to a threshold of $\frac{1}{2}$ the RMS noise. The 15 GHz observations achieve a sensitivity of $7~\rm \mu Jy~beam^{-1}$. The size of the resulting synthesized beam is $1\farcs88\times 0\farcs42$ ($\sim 180$ au) at a position angle of $10.5^{\circ}$. Due to the large beam, the emission is only marginally resolved along the minor axis of the beam. The peak flux is $0.044~\rm  mJy~beam^{-1}$ as measured in the CLEANed image. We used the {\scriptsize CASA} task \textit{uvmodelfit} and a disk model to obtain a total flux of $0.050\pm0.008~\rm  mJy$.

The 45 GHz VLA data were acquired on 2019 May 13 and 2019 July 10 and had a total combined on-source time of $2345\rm~s$. Both 45 GHz SBs used a Q Band tuning setup with $4\times2.048$ GHz basebands and rest frequency centers of 41 GHz, 43 GHz, 47 GHz, and 49 GHz. This gives an effective frequency of 45 GHz (6.7 mm) for the Q band. The 45 GHz observations were concatenated in {\scriptsize CASA} and the data were imaged with a natural weighting and cleaned using the \textit{CLEAN} algorithm down to a threshold of $\frac{1}{2}$ the RMS noise. Together, these data achieve a sensitivity of $43~\rm \mu Jy~beam^{-1}$. The resulting synthesized beam is $0\farcs36 \times 0\farcs 16$ ($\sim 40$ au) at a position angle of $10.2^{\circ}$.

The low declination of EX Lupi leads to an elongated synthesized beam with the VLA (EX Lupi is located at $\delta = -40^{\circ}$ and the VLA is at a latitude of $34^{\circ}~ \rm N$). The atmospheric fluctuations at low altitude can also be much greater, leading to a poorer phase calibration and thus impacting the quality of the reconstructed images. These factors, coupled with less time on source than initially requested for the Q Band, led to a non-detection at 45 GHz. We note that this does not necessarily mean the disk is not observable at these frequencies. 

\subsection{ALMA Observations}

EX Lupi was observed with ALMA (2017.1.00224.S, PI K\'osp\'al) on 2018 January 27, March 17, and March 19 with a phase center of RA$ = 16^{\rm h}~ 03^{\rm m} ~ 5.5^{s}$  $\delta = -40^{\circ}$ $18'$ $25\farcs4$. The total on-source integration time was 11640 s. The nominal antenna configuration was C43-4, with baselines between 14 m and 1398 m. Two spectral setups were used with spectral windows for different molecular lines (which will be presented in a later paper). In both setups, a 1.875 GHz wide window was centered at 100.2 GHz (2.99 mm) to measure the continuum emission of EX Lupi. The data were manually calibrated using the {\scriptsize CASA 5.4.1} pipeline. The procedure included offline water vapor radiometer calibration, system temperature correction, and bandpass, phase and amplitude calibrations. Quasars J1427$-$4206 and J1517$-$2422 were used for pointing, bandpass, and flux calibration, and Quasar J1610$-$3958 was used for phase calibration. The sampled visibilities were Fourier transformed, creating dirty images of the source, which yields the skymodel of the target convolved with the point source function (PSF) of the beam. Dirty images for each spectral window were used to determine the frequencies without line emission and as input for the \emph{uvcontsub} routine in {\scriptsize CASA} to obtain the continuum emission.

The 100 GHz observations were concatenated in {\scriptsize CASA} and the data were imaged with a natural weighting and CLEANed using the \textit{CLEAN} algorithm down to a threshold of $\frac{1}{2}$ the RMS noise. Together, these data achieve a sensitivity of $13~\rm \mu Jy~beam^{-1}$. The resulting synthesized beam is $1\farcs09 \times 0\farcs 81$ ($\sim 150$ au) at a position angle of $87.8^{\circ}$. The disk is marginally resolved. The peak flux is $1.98~\rm Jy~beam^{-1}$ as measured in the CLEANed image. We used the {\scriptsize CASA} task \textit{uvmodelfit} and a disk model to obtain a total flux of $2.72\pm0.013~\rm \mu Jy$.

\subsection{Literature Data}

In addition to the new ALMA and VLA data presented here, we also use the ALMA 232 GHz continuum data from \citet{hales18} in our analysis. The observations were made on 2016 July 25, have a CLEANed synthesized beam of $0\farcs32\times0\farcs26$, and sensitivity of $\sigma_{\rm rms} = 38 ~\rm \mu Jy~beam^{-1}$. The disk was well resolved at this frequency. No further calibration or processing was performed outside the procedure listed in \citet{hales18}.

\section{Model Fitting}\label{sec:models}

To constrain the parameters of the dust in EX Lupi's circumstellar disk, we followed a radiative transfer (RT) model fitting approach similar to \citet{hales18}. We use the RT code {\scriptsize RADMC-3D 0.41} \citep{dullemond12} with the Python interface radmc3dPy\footnote{http://www.ast.cam.ac.uk/~juhasz/radmc3dPyDoc/index.html} to set the code input parameters for a given disk model. We keep the following parameters fixed in the fitting procedure: inclination $i=32^{\circ}$, position angle PA$=65^{\circ}$, and inner disk radius $\rm r=0.05~au$ \citep{hales18}; flaring parameter $\psi = 0.09$ \citep{sipos09}; stellar temperature T $=3859~\rm K$ and stellar radius R $=1.6~\rm R_{\odot}$ \citep{frasca17}. 

We adopt a disk model similar to that of a typical T Tauri protoplanetary disk \citep{andrews09} with a disk density given by:
\begin{equation}
\rho =\frac{\Sigma(r,\phi)}{H_p\sqrt{(2\pi)}}\exp{\left(-\frac{z^2}{2H_p^2}\right)},
\end{equation}
where $\Sigma$ is the surface density profile, $H_{\rm p}$ is the pressure scale height, and z is the height above the disk midplane. The disk's surface density profile follow a power-law profile with an exponential outer tapering:
\begin{equation}
\Sigma(r) = \Sigma_0\left(\frac{r}{R_c}\right)^{-\gamma} \exp{\left\{-\left(\frac{r}{R_c} \right)^{2-\gamma}\right\}},
\end{equation}
where $\Sigma_{0}$ is the surface density at the inner radius of 0.05 au, $R_{\rm c}$ is the characteristic radius of the disk, and $\gamma$ is the power-law exponent of the radial surface density profile. The pressure scale height is defined as:
\begin{equation}
H_p = h_c \left( \frac{r}{100 \rm~au} \right) ^{1+\psi}, 
\end{equation}
where $\psi$ is the disk flaring parameter and $h_{\rm c}$ is the ratio of the pressure scale height over radius at 100 au \citep[see][]{hales18}. 

In order to converge on the best fit disk parameters, we use a Metropolis-Hastings Markov Chain Monte Carlo (MCMC) model fitting approach. The free parameters considered in the modeling are: the total disk mass with a gas-to-dust ratio of 100:1, $M_{\rm disk}$, characteristic radius, $R_{\rm c}$, power law exponent of the surface density profile, $\gamma$, and scale height ratio at 100 au, $h_{\rm c}$. We perform the MCMC modelling in the image plane \citep[see, e.g., ][]{booth16,white_fom}. After a trial model is selected, we use {\scriptsize RADMC-3D} and the \textit{mctherm} command to calculate the dust temperature and then use the \textit{image} command to generate a ray-traced continuum image projected to the inclination and position angle of the disk. The image is then attenuated by the primary beam and convolved with the dirty beam for a given observational setup. A $\chi^{2}$ for each trial model is calculated as
\begin{equation}
    \chi^{2} = \frac{(Data - Model)^{2}}{\sigma^{2}},
\end{equation}
where $\sigma$ is the observed $\sigma_{\rm rms}$ for a given observation multiplied by the synthetic beam size in pixels \citep[see ][]{booth16}. To fit multiple wavelength's data simultaneously, the $\chi^{2}$ at each wavelength needs to be calculated and weighted. We adopt an equal weighting for each wavelength, and all of the $\chi^{2}$ values are averaged together. A given trial model is then accepted if a random number drawn from a uniform distribution [0,1] is less than $\alpha$, where 
\begin{equation}
    \alpha = \min(e^{\frac{1}{2}(\chi^{2}_{i} - \chi^{2}_{i+1})},1).
\end{equation}

Fitting the 232 GHz, 100 GHz, and 15 GHz data simultaneously requires a dust opacity file extended to larger grain sizes. We use the {\scriptsize OpacityTool}\footnote{The OpacityTool Software was obtained from https://dianaproject.wp.st-andrews.ac.uk/data-results-downloads/fortran-package/} program \citep{toon81, woitke16} to get more realistic dust absorption and scattering parameters. This program calculates dust opacities by using a volume mixture of 60\% amorphous silicates \citep[e.g.,][]{dorschner95}, 15\% amorphous carbon \citep[e.g.][]{zubko96}, and a 25\% porosity. Bruggeman mixing is used to calculate an effective refractory index and a distribution of hollow spheres with a maximum hollow ratio of 0.8 \citep{min05} is included to avoid Mie theory scattering artifacts. We further assume the disk is populated by $0.1 - 21000~\mu$m grains following a power-law size distribution of $s^{-3.5}$. 

We were unable to adequately reproduce any wavelength's data with this approach. We tried varying the weight of each wavelength's $\chi^{2}$ and could only begin obtaining reasonable results if the weighting for the 15 GHz data was set to an arbitrarily small value. This indicates that the approach is not well suited for fitting all three datasets simultaneously, or that there are different emission mechanisms at longer wavelengths. Therefore, we decided to exclude the 15 GHz data and discuss other approaches to fitting it in Sec.\ref{sec:mod}.

\begin{table*}
\begin{center}
\caption{Summary of the most probable model parameters and the corresponding 95\% confidence intervals from the posterior distributions in brackets. The first row is the ALMA 232 GHz and 100 GHz data fit simultaneously, the second row is the 15 GHz VLA data fit alone, the third row is fitting a point source model on top of the RT calculated 15 GHz disk emission. The disk mass is the total disk mass assuming a gas-to-dust ratio of 100:1. The scale height is measured at 100 au \citep[see][]{hales18}. }\label{mod_results}
\bigskip
\begin{tabular}{c | c c c c }
\hline

Dataset(s) & Disk Mass & Characteristic Radius & $\gamma$ & Scale Height \\
 & (M$_{\odot}$) & (au) &  & (au) \\
 \hline\hline
232 GHz + 100 GHz & 0.0099  & 45 & 0.25 & 2.5 \\
 & [0.0067, 0.012] & [39, 68] & [$-0.36$, 0.77] & [1.9, 3.8] \\
15 GHz & 0.035 & 56 & 0.12 & 89 \\
 & [0.011, 0.053] & [13, 140] & [$-1.8$, 1.2] & [21, 98] \\
 
\\

\hline
Dataset & Flux & X-offset & Y-offset & \\
 & (mJy) & ($''$) & ($''$) & \\
\hline\hline
15 GHz & 0.055 & 0.025 & 0.26 & \\
  & [0.045, 0.066] & [$-0.09$, 0.11] & [$-0.010$, 0.55] & \\

\end{tabular}
\end{center}
\end{table*}

\begin{figure*}
\centering
\includegraphics[width=\textwidth]{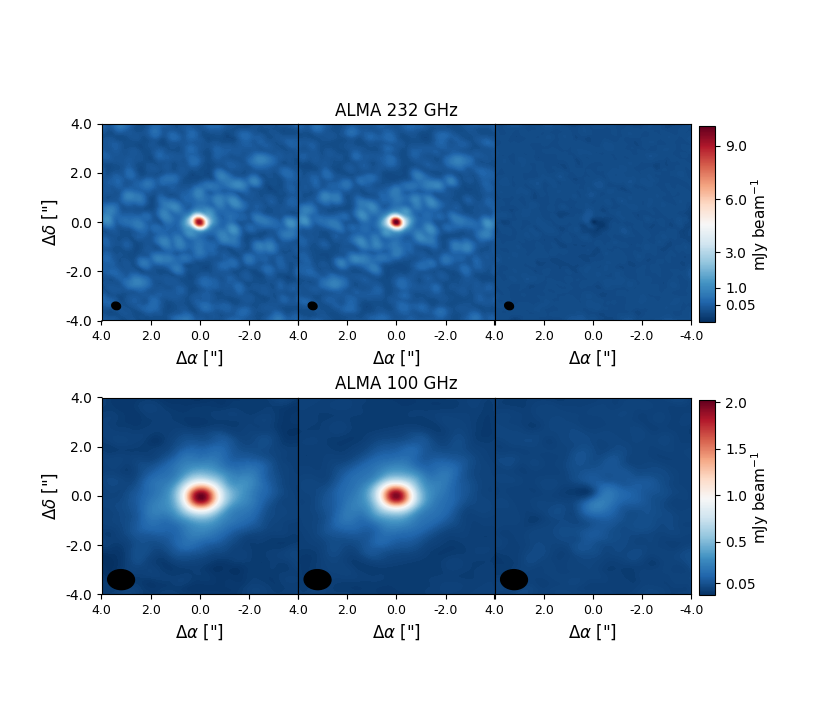}
\caption{\textbf{Top:} Dirty image, best fit RT model, and residuals for the ALMA 232 GHz observations. \textbf{Bottom:} Dirty image, best fit RT model, and residuals for the ALMA 100 GHz observations. The two datasets were fit simultaneously with equal weighting. The black ellipses in the bottom left of each figure represent the synthesized beam.  \label{fig:resid_alma}}
\end{figure*}

\begin{figure*}
\centering
\includegraphics[width=0.75\textwidth]{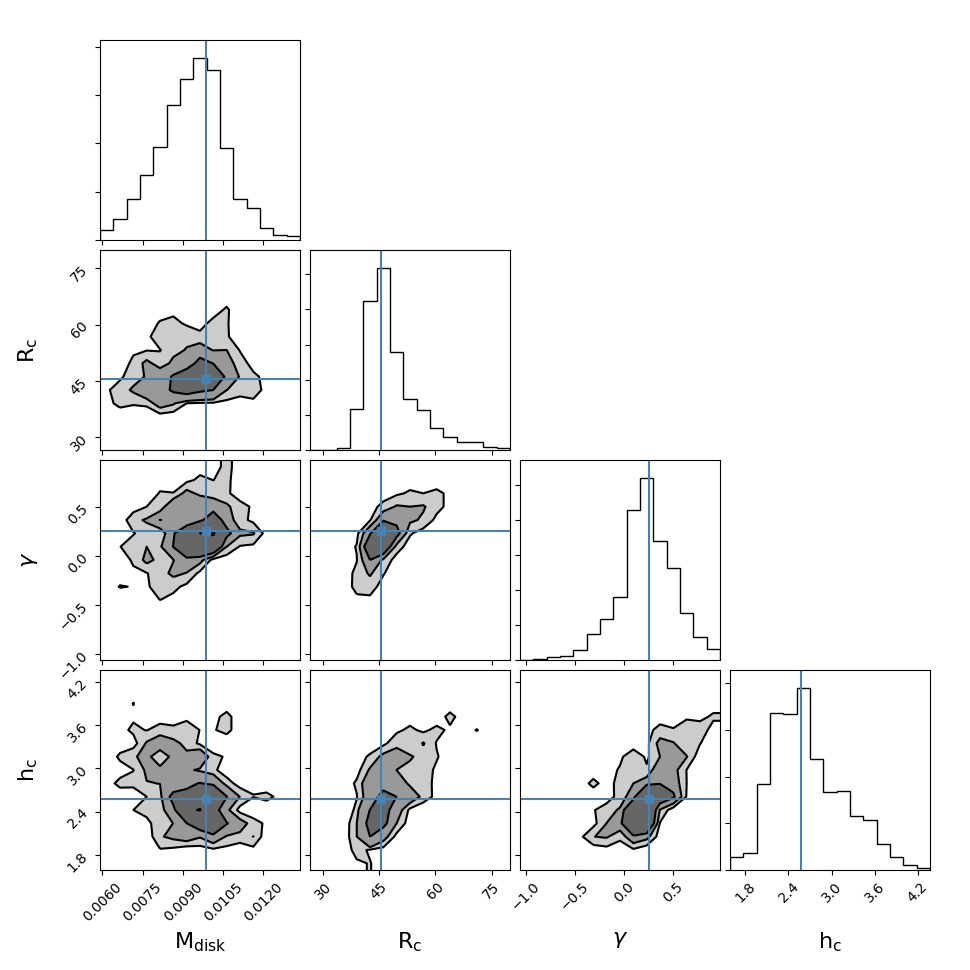}
\caption{Posterior distribution of the best fit model for fitting the ALMA 232 GHz and 100 GHz simultaneously. The most probable values for each parameter are denoted by the blue lines. \label{fig:pdf_alma}}
\end{figure*}

To constrain the properties of the 232 GHz and 100 GHz observations, we use all the same approach and assumptions outlined above but change the particle size population to $0.1 - 10000~\mu$m grains. We adopted an equal weighting for the two data sets and ran an MCMC fit with $100\times1000$ link chains minus 100 each for burn-in. This approach of only fitting the ALMA data was able to well reproduce the observations. The most probable values of the free parameters and 95\% credible region (CR) are summarized in Table\,\ref{mod_results}. The resulting best fit model and residuals are shown in Fig.\,\ref{fig:resid_alma} and the posterior distribution functions (PDF) are shown in Fig.\,\ref{fig:pdf_alma}.

\section{Discussion}\label{sec:disc}

\subsection{Model Fitting Results}\label{sec:mod}

Our RT models of the 232 and 100 GHz ALMA data were able to well reproduce the observations. When fitting both of the datasets simultaneously, we find most probable values of: M$_{\rm disk} = 0.01~\rm M_{\odot}$, R$_{c}=45$ au, $\gamma= 0.25$, and h$_{c} = 2.5$ au (the corresponding 95\% confidence intervals are listed in Table\,\ref{mod_results}). The total flux of the most probable models at 232 and 100 GHz are 18.2 and 2.45 mJy, respectively. Most of the results are consistent with the values reported in \citet{hales18}, within the uncertainties, which only fit the ALMA 232 GHz data. We find though that the most probable value of the characteristic radius is about a factor of 2 larger. This difference in radius could be due to the larger beam size in the 100 GHz observations, along with equal weighting, which is forcing the models to be larger. The spectral index between 232 and 100 GHz is $\alpha_{1.3-3.0 {\rm mm}} = 2.20\pm 0.11$ (the uncertainty includes a 10\% absolute flux calibration uncertainty at each frequency). The spectral index is consistent with $2.19\pm0.47$ as reported in \citet{hales18} and calculated within the ALMA Band 6 spectral windows. Extrapolating the peak flux of the 100 GHz data to 45 GHz, along with the new beam size and $\alpha=2.2$, gives a peak flux lower than the achieved $\sigma_{\rm rms}$ of the 45 GHz observations. This calculated flux shows that the non-detection at 45 GHz is not useful for constraining the thermal disk properties. 

\begin{figure*}
\centering
\includegraphics[width=\textwidth]{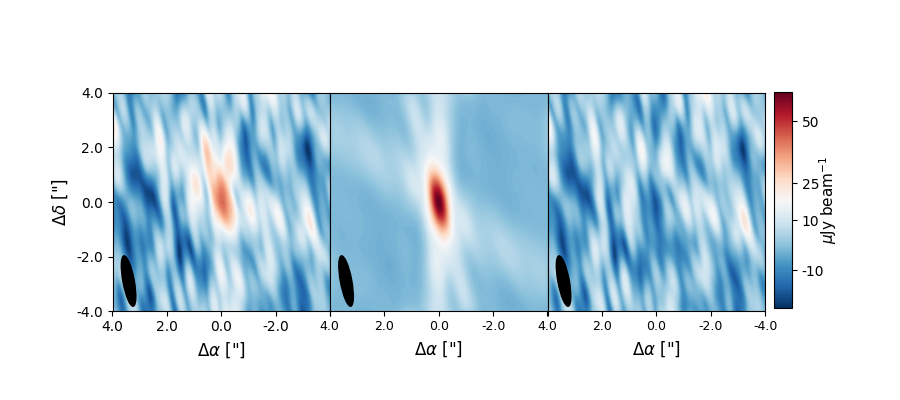}
\caption{\textbf{Left:} VLA 15 GHz dirty image. \textbf{Middle:} 15 GHz RT disk model \textbf{Right:} Data minus model residuals. The black ellipses in the bottom left of each figure represent the synthesized beam.  \label{fig:resid_vla}}
\end{figure*}

\begin{figure*}
\centering
\includegraphics[width=0.75\textwidth]{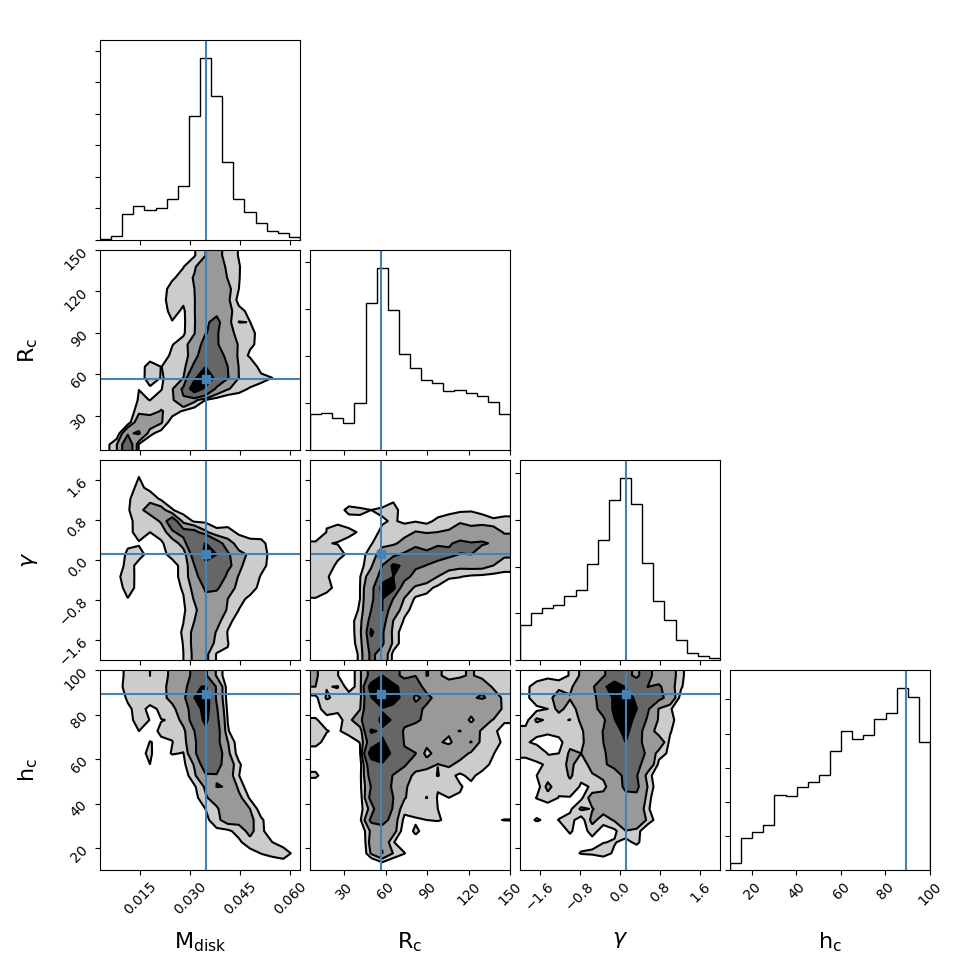}
\caption{Posterior distribution of the best fit model of the 15 GHz emission alone shown in Fig.\,\ref{fig:resid_vla}. The most probable values are denoted by the blue lines. \label{fig:pdf_vla}}
\end{figure*}

If the 15 GHz observations are tracing the thermal emission from large grains, then it is possible these large grains are located at a different area of the disk than the smaller grains probed by ALMA. To test this, we tried fitting the VLA 15 GHz data alone with the same RT approach as for the ALMA datasets. This approach allows for a different disk geometry of the large grains, but still requires the total flux to be well fit to the data. We note though that due to the large beam size at 15 GHz, the disk is not resolved along the major axis and would be only marginally resolved along the minor axis.  The results are summarized in Table\,\ref{mod_results}, the best fit model and residuals are shown in Fig.\,\ref{fig:resid_vla}, and the PDF is shown in Fig.\,\ref{fig:pdf_vla}.

While the 15 GHz only model does indeed well reproduce the data, the resulting disk parameters seem improbable.  As segregation by grain size is possible due to radial drift and settling, a different disk geometry alone is not immediately disqualifying. The scale height, however, seems un-physically large at 89 au. Larger grains would be expected to settle in the midplane of the disk, meaning the scale height would likely be the same or smaller than is observed with the mm grains. The best fit total disk mass is also $3.5\times$ larger than when fitting to the mm data alone. Such a disk may have emission at 45 GHz, depending on the spectral index, and we report a non-detection at 45 GHz.  We conclude that even though a thermal emission model can technically fit the data well at 15 GHz, the resulting disk parameters to do so are highly improbable. Alternate sources of 15 GHz emission are explored in Sec.\,\ref{sec:long}.

\subsection{Millimeter Observations \& Grain Growth}

EX Lupi lies between the Lupus 3 and Lupus 4 star forming regions \citep{cambresy99}. ALMA surveys of protoplanetary disks in the Lupus star forming regions have found total disk masses to be $\sim10^{-3}~\rm M_{\odot}$ \citep{ansdell16}. Our model fitting finds a total disk mass nearly an order of magnitude higher, assuming a gas-to-dust ratio of 100:1. The mass discrepancy could be due to the assumptions made in the mass calculation in \citet{ansdell16} such as the optical depth and that here we include a full RT calculation. The actual gas-to-dust ratio in EX Lupi could also be much lower than assumed. If the difference in the masses is indeed real, it could be explained by EX Lupi being younger, more heavily accreting, or from an inherent difference in the disks of EXor/FUor-type system from that of typical protoplanetary disks. EX Lupi's disk mass is smaller than that of FUors, as expected, but the characteristic radius is similar. Other ALMA studies have found that EXor/FUor disks tend to be more compact than that of typical protoplanetary disks \citep[e.g.,][K\'osp\'al et al. in prep.]{cieza18}. 

EX Lupi experienced an outburst in 2008 and has since returned to a quiescent state. Therefore, all of the ALMA and VLA data presented here are indicative of a post-outburst circumstellar environment. \citet{abraham09} found that the outburst increased the crystallinity of the disk grains and transported them from the inner regions to the outer regions. Their work shows that even a short-lived outburst, such as an EXor-type outburst, can have observable effects on the circumstellar disk.

The water ice line (or snow line) is the radial location in a disk where water reaches its condensation temperature and freezes out on to grains in the disk. The exact temperature at which this occurs depends on other disk properties, such as gas pressure, but is typically between 150 to 175 K in the midplane of a disk \citep{lecar06}. EX Lupi's recent outburst increased the bolometric luminosity by about $36 \times$ \citep{abraham09}. This enhanced luminosity can heat the disk and push the ice line(s) out to further radial distances. This was observed in ALMA observations of V883 Ori, where the ice line moved from a presumed $1-5$ au pre-outburst to a current location of 42 au as measured on the surface of the disk \citep{cieza16}. As an outburst fades and the temperature drops, the ice line will move back inwards to the pre-outburst location. From \citet[][Fig.\,4 in supplementary materials]{abraham09}, we estimate that during EX Lupi's outburst the ice line was located at $20-30$ au on the surface of the disk and $1-2$ au in the midplane of the disk. Using \textit{Spitzer} observations of the H$_{2}$O spectra, \citet{banzatti12} find the ice line moved from 1.3 au during outburst to 0.6 au after outburst. 

\begin{figure}
\centering
\includegraphics[width=0.51\textwidth]{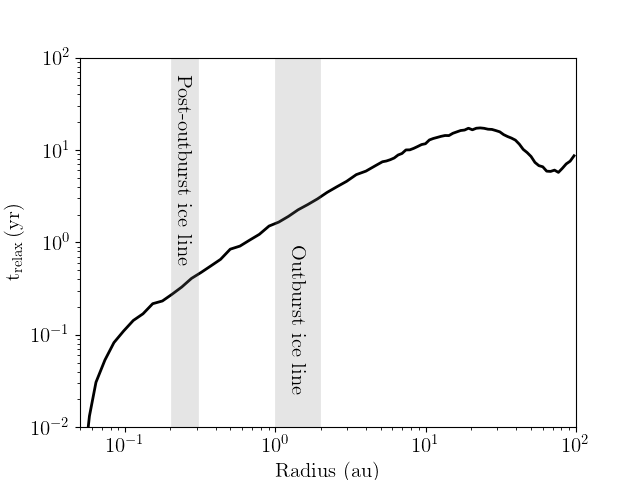}
\caption{ The relaxation time in years as a function of disk radii along the midplane for the best fit RT model of EX Lupi \citep[see Eqns.\,10-13 in][]{flock17}.  The locations of the ice line along the disk midplane both during \citep{abraham09} and post-outburst are also denoted in gray. \label{fig:time_plot}}
\end{figure}

Our RT modeling shows that the ice line is now located at $3-4$ au on the surface of the disk and $0.2-0.3$ au in the disk's midplane, indicating a significant shift inwards post-outburst. To check if the disk could have reasonably cooled off enough to make this change in the position of the ice line, we explore the disk relaxation timescales \citep[see \citet{lin15} and Section 2.3 of][]{flock17}. The density and temperature we take from our best-fit RT model, the gas-to-dust ratio is set to 100:1, and we choose a representative cooling wavelength opacity at $\rm 18\,\mu m$ of $\kappa=52.7 \rm \,cm^2/g$. The relaxation time curve is shown in Fig.\,\ref{fig:time_plot}. 

The black line represents the relaxation timescale in years for the best fit disk model as a function of radial position in the disk, with the outburst and post-outburst ice lines indicated in gray. This curve shows that EX Lupi's recent outburst, of duration of $\sim1$ year, could only significantly heat the midplane of the disk out to radii of 1-2 au, consistent with data from during the outburst \citep{abraham09,banzatti12}. While the current data cannot be used to infer the location of the ice line prior to the outburst, the outburst and post-outburst data together show evidence that the location of the ice line can indeed shift during EXor-type events. The plot also shows that the $\sim10$ yr time between the end of the outburst and the observations is more than enough to relax the disk up to 10 au to a new thermal state.

An ice line can be a source of turbulence in a disk, around which grain growth can occur \citep[e.g.,][]{brauer08, ros13, zhang15}. Therefore as the ice line moves inwards after an outburst it provides a potential mechanism to spur grain growth throughout the disk. The associated timescales for water to deposit onto grains near the ice line will be much shorter than the relaxation timescale outlined above, meaning grain growth from water freeze-out or deposition can be commensurate with the moving position of the ice line  \citep[e.g.,][]{brown90}. Thermal grains are inefficient emitters at wavelengths longer than their grain size. Therefore, using the detection of the disk in the ALMA 100 GHz data, and calculated spectral index of $\alpha=2.20\pm0.01$ indicating the disk is optically thin, we infer that grains of at least $3\rm~mm$ may be present in the disk post-outburst. Our RT models include grains of sizes up to $1\rm~cm$. While this is not a confirmation of $1\rm~cm$ grains it does indicate the data is at least consistent with the presence of up to $1\rm~cm$ grains. Significant grain growth is possible with EX Lupi's evolving ice line, although we note that there is no pre-outburst ALMA and VLA data to compare it to. If EXor-like outbursts, which are short lived and can repeat many times during pre-main-sequence stellar evolution \citep[e.g.,][]{vorobyov15, white19}, are indeed common for most stars then appreciable grain growth can occur throughout the disk both early and often. 
Quick and abundant grain growth to mm-cm sized particles is an important step in the planet formation process and can drive the growth of larger planets and planetesimals \citep[e.g.,][]{morbidelli16, johansen17, hughes17}.
Significant early grain growth and the presence of planetesimals is also a possible explanation for the gaps seen in HL Tau \citep{brogan15} or surveys of nearby protoplanetary disks with the ALMA DSHARP survey \citep[e.g.,][]{andrews18}. 

\subsection{Long Wavelength Central Emission}\label{sec:long}

\begin{figure*}
\centering
\includegraphics[width=\textwidth]{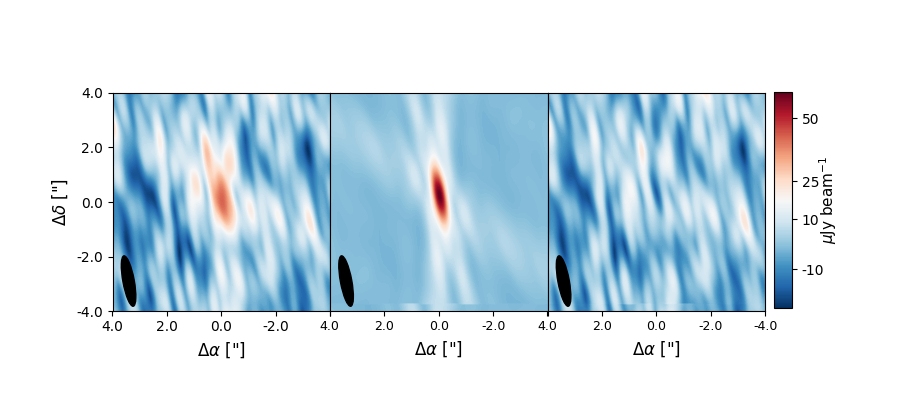}
\caption{\textbf{Left:} VLA 15 GHz dirty image. \textbf{Middle:} 15 GHz RT disk model (as calculated from fitting to both ALMA datasets) added to the best fit central emission model. \textbf{Right:} Data minus model residuals. The black ellipses in the bottom right represents the synthesized beam.  \label{fig:resid_cent}}
\end{figure*}

\begin{figure*}
\centering
\includegraphics[width=0.75\textwidth]{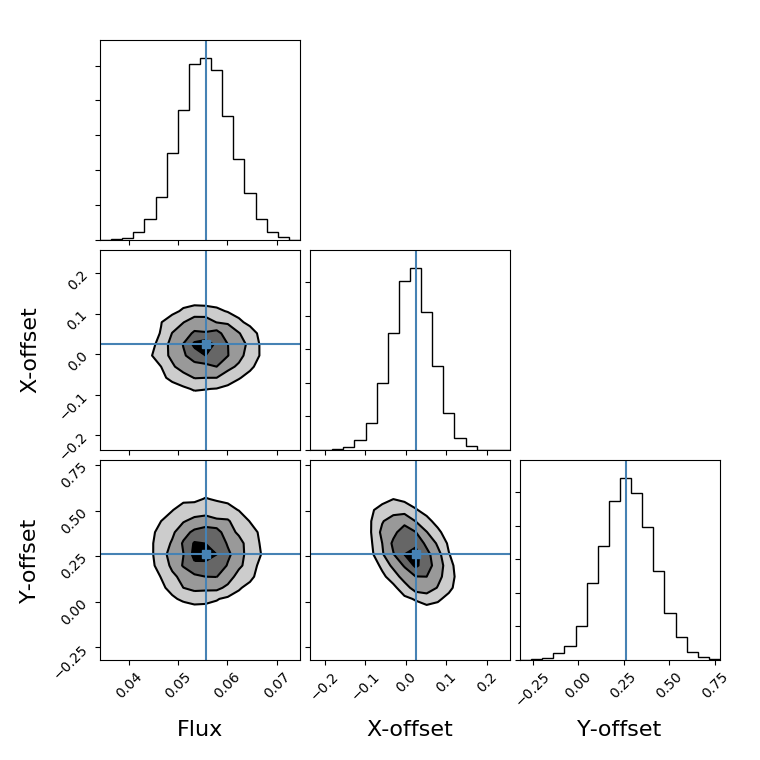}
\caption{Posterior distribution of the 15 GHz central emission on top of the RT calculated disk emission from the ALMA data. The most probable values are denoted by the blue lines. \label{fig:pdf_cent}}
\end{figure*}

In Sec.\,\ref{sec:models}, we concluded that the 15 GHz VLA data is likely dominated by something other than thermal disk emission. A more likely scenario is that at 15 GHz we are seeing thermal disk emission plus a combination of non-thermal disk emission, stellar winds, jets, or stellar emission. To test this scenario, we can fit a point source emission model on top of the thermal disk emission. Given the resolution of the 15 GHz data ($1\farcs88\times0\farcs42$), we are unable to differentiate emission from a point source and a moderately extended (i.e., a few au) region.  Taking the best fit RT model from the ALMA data alone, we use RADMC-3D to calculate the emission at 15 GHz, resulting in a very faint disk with a total flux of only $\sim0.001$ mJy. Starting with this disk model, the ``additional" emission required to reproduce the data can be assumed to be coming from approximately the center of the disk. This leaves only 3 free parameters: the flux of the central emission, the X-offset, and the Y-offset. We adopt an MCMC modelling approach similar to the one outlined in Sec.\,\ref{sec:models} but now just add the flux of the central emission on top of the already calculated RT model. The results are summarized in Table\,\ref{mod_results}, the best fit model and residuals are shown in Fig.\,\ref{fig:resid_cent}, and the PDF is shown in Fig.\,\ref{fig:pdf_cent}. We find a most probable flux of 0.055 mJy, an X-offset of $0\farcs025$, and a Y-offset of $0\farcs26$. Due to the nature of the large beam at 15 GHz ($1\farcs88\times 0\farcs42$), the location of the central emission flux is still broadly consistent with being peaked on the star itself.

\subsubsection{Role of the Magnetic Field}\label{sec:mag_field}

Some of the most promising outburst theories for EXors rely on the role of the magnetic field. \citet{armitage16} predicted that the outbursts could be explained by changes in the polarity and strength of the stellar magnetic fields at the kG-level. In a competing scenario, \citet{dangelo12} proposed an instability which can lead to quasi-periodic oscillations in the inner disk and associated recurrent outbursts. This instability can occur when the accretion disk is truncated close to the co-rotation radius by the strong magnetic field of the star. Stable accretion columns linked to a very strong magnetic field in EX Lupi were noted in \citet{aguilar15}.

In principle, if kG-level stellar magnetic fields were present in EX Lupi then there could be significant non-thermal emission that would yield brightness temperatures easily detectable at long wavelengths with VLA. The central emission flux of 0.055 mJy corresponds to a Rayleigh-Jeans brightness temperature of T$_{B} \approx 4\times10^{7}$ K, assuming the size of the emitting region is uniformly spread out over the surface of the star (we note that depending on the emission mechanism the actual emitting region could range from a small localized area of the star to several stellar radii). This brightness temperature is much larger than the effective temperature of EX Lupi ($T=3859~\rm K$), which indicates that significant non-thermal stellar emission could be present as was seen in optical line emission \citep[e.g.,][]{aguilar12}.

One potential source of such large brightness temperatures is synchrotron emission from relativistic or nearly relativistic electrons being accelerated by EX Lupi's magnetic field. If synchrotron emission is present, then the flux should peak near the critical frequency \citep[e.g.,][]{hughes19},
\begin{equation}
\nu_{crit} = \frac{\gamma^{2} q B} {2 \pi m_{e}},
\end{equation}
where $\gamma$ is the Lorentz factor and is assumed to be $\sim1$, q is the electron charge, B is the magnetic field strength, and $m_{e}$ is the electron mass (we note that $\gamma$ here is not the same as the disk power-law exponent used in Eqn.\,2). Adopting a magnetic field strength of 3 kG gives $\nu_{crit} = 8.4$ GHz. Since the flux should quickly drop off at frequencies larger than the critical frequency, it would be unlikely to observe such a large flux at 15 GHz. However, if the electrons are slightly more relativistic, with $\gamma=1.34$, then $\nu_{crit}=15$ GHz. Alternatively, \citet{armitage16} predict that significant changes in magnetic field strength could be a driver for the episodic accretion in EXor-type stars. Therefore, the magnetic field strength could potentially be different than measured previously with CFHT \citep{kospal20}. If this is indeed the case, and 
$\nu_{crit}$ is $\sim15$ GHz, then a magnetic field strength of $\sim5.4$ kG can be inferred.

The effective temperature \citep[see, e.g.,][]{pacholczyk70} of electrons emitting synchrotron radiation at a frequency $\nu$, is given by:
\begin{equation}
T_{e}\approx \Big(\frac{2 \pi m_{e} \nu }{q B}\Big)^{1/2} \frac{m_e c^2}{3 k}
\end{equation}
where k is the Boltzmann constant. This yields  $2-3 \times 10^{9}$ K, depending on which value for the magnetic field is used. The electron temperature is about a factor of 50 larger than the observed brightness temperature implying the emission is optically thin. At face value, synchrotron emission is a potential source of the 15 GHz emission. An important consideration though is that the size of the emitting region could be extended much further from the stellar surface. At further stellar separations, the magnetic field strength will become weaker causing $\nu_{crit}$ to fall below 15 GHz and the expected synchrotron emission at 15 GHz to be much lower as well.

While EX Lupi is indeed still a pre-main-sequence star, its spectral classification is M0 and its mass indicates that when it reaches the main-sequence it will still be an M-type star. M-type stars are notorious for flares and magnetic activity, but were not observed at radio wavelengths until recent technological advancements \citep{berger01}. While the exact emission mechanisms are still debated, they are thought to be primarily due to electron cyclotron maser instabilities (ECMI) and/or gyrosynchrotron emission. 

ECMI emission is similar to the auroral emission observed on most solar system planets \citep{turnpenney18}. If the 15 GHz emission in EX Lupi was due to ECMI, then the emission is expected to peak at the fundamental critical frequency of $\nu_{crit} = 8.4$ GHz and fall off rapidly at higher frequencies. ECMI requires a relatively stable magnetic field configuration, such as in planetary or brown dwarf magnetic fields. Therefore, even though EX Lupi's magnetic field strength could have changed between the CFHT and VLA observations such that $\nu_{crit} \sim 15$ GHz, ECMI is a highly unlikely source of the 15 GHz emission due to the required stability. 

\begin{figure}
\centering
\includegraphics[width=0.51\textwidth]{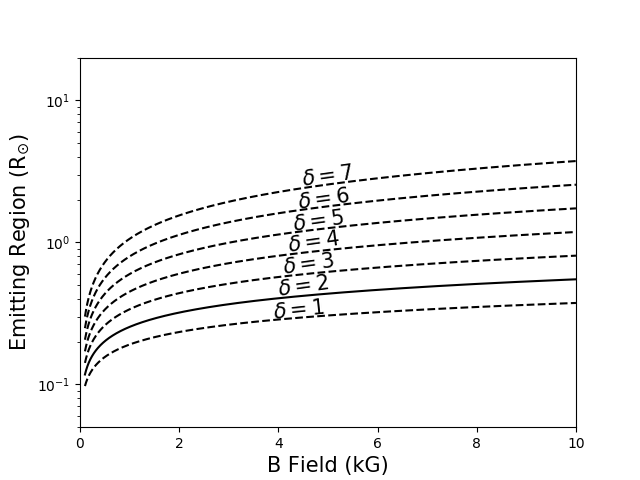}
\caption{ The size of the emitting region and magnetic field strength for various electron energy indices, $\delta$, assuming gyrosynchrotron emission at 15 GHz. The solid black line at $\delta=2$ represents a typical lower level estimate for $\delta$. \label{fig:gyro_plot}}
\end{figure}

If the 15 GHz emission in EX Lupi was due to gyrosynctrotron emission, then it is likely due to magnetic reconnection events releasing a large number of energetic particles \citep[e.g.,][]{williams14, hughes19}. The surface of a pre-main-sequence accreting M-type star, such as EX Lupi, is undoubtedly a turbulent environment where these processes could dominate. The expected emission at 15 GHz depends on the size of the emitting region (which is typically much smaller than the stellar radius), the magnetic field strength, and the electron energy index $\delta$. Assuming the gyrosynchrotron emission is optically thin, then the spectral index can be used as a tracer for $\delta$. Following \citet{hughes19}, the relation between all of these parameters is given by:
\begin{equation} 
    R_{\rm em} = 132 \Big(\frac{d}{pc}\Big) \Big(\frac{GHz}{\nu}\Big) \sqrt{\frac{F_{\mu Jy}}{T_{eff}}}, 
\end{equation}

where $R_{\rm em}$ is the emitting region in units of R$_{\odot}$, F$_{\rm  \mu Jy}$ is the observed flux in $\mu$Jy, and T$_{\rm eff}$ is given by:

\begin{equation}
    T_{\rm eff} = 2.2\times 10^9 ~ ({\rm sin}~\theta)^{-0.36 - 0.06\delta} 10^{-0.31 \delta}  \Big(\frac{\nu}{\nu_{crit}}\Big)^{0.50 + 0.085 \delta}
\end{equation}

where $\theta$ is assumed to be $\sim90^{\circ}$ \citep[see][for the derivation and further details on the relationship between expected flux and the size of the emitting region]{dulk85}. Fig.\,\ref{fig:gyro_plot} shows the size of the emitting region and magnetic field strength for various values of $\delta$ (which cannot be constrained with the available data). If we assume a typical lower level limit of $\delta\sim2$ and a magnetic field strength range of $3-6~\rm kG$ then the corresponding size of the emitting region is $\sim0.3-0.5~\rm R_{\odot}$. This is a significant fraction of the stellar surface ($\rm R_{*}=1.6~ R_{\odot}$) and much larger than is typically expected for main-sequence M-dwarf stars which have reconnection regions typically of order $0.01~\rm R_{\odot}$. If $\delta>2$, which is typically the case in M-dwarfs, then the size of the emitting region quickly becomes larger than the star. EX Lupi could, however, have a very small value of $\delta$ or have the gyrosynchrotron emission come from a large area where it is actively accreting disk material. Follow-up observations at higher/lower frequencies will enable spectral index constraints, and thus a value for $\delta$, to better determine if gyrosynchrotron is a potential source of the observed 15 GHz emission.

\subsubsection{Radio Jets and Non-thermal Disk Emission}\label{sec:jets}

Some EXor/FUor systems also have very bright radio jets \citep[e.g., Z CMa and L1551 IRS 5,][]{poetzel89,rodriguez03}. While EX Lupi did show significant X-ray activity leading up to and during the most recent outburst, the emission is most likely stemming from accretion shocks instead of jets \citep[see, e.g.,][]{grosso10}. No radio jets have been previously reported in EX Lupi. If the 15 GHz emission was indeed due to a radio jet, then we would expect it to have an approximately flat spectral index leading it to possibly be detectable at 45 GHz. Since we did not detect anything at 45 GHz, and no radio jet was reported previously, we note the likelihood of the 15 GHz emission being jet-driven is low. 

Centrally located disk winds are another possible source of long wavelength emission. \citet{kospal11} found that the hydrogen emission within 1 au is likely due to disk winds in EX Lupi. In optical line spectra, there is further evidence of disk winds in both outburst and quiescence \citep{aguilar12, aguilar15, banzatti19}. \citet{hales18} observed large scale asymmetries in the outer CO gas disk which could be due to a molecular outflow. Similar to jets, disk winds would likely be peaked at frequencies $<15$ GHz and have a flat to slightly positive spectral index. Follow-up observations at $<15$ GHz are necessary in order to measure the spectral index and determine if winds or jets are a possible source of observed emission. 

Free-free emission from an ionized disk, typically more prevalent in the latter stages of disk dissipation when photoevaporation becomes significant, can come from EUV and X-ray irradiation. EX Lupi had significant X-ray activity before and during the outburst, which could have significantly ionized its inner disk. Although unlikely, if this irradiation continued after the outburst ended then free-free emission could still be present in the EX Lupi system. Free-free emission can also arise from accretion shocks propagating through the disk when the accretion rate becomes much larger \citep[e.g.,][]{hartmann96}. \citet{aguilar15} find evidence of accretion shocks via optical spectroscopy and \citet{grosso10} noted that the X-ray emission seen in EX Lupi is also likely due to accretion shocks. The X-ray luminosity of EX Lupi was $1.7\times10^{30}~erg~s^{-1}$ during the outburst \citep{grosso10}. Using the relation for X-ray luminosity to expected radio flux from free-free emission outlined in \citet{pascucci12}:
\begin{equation}
    F_{3.5 \,cm} = 2.4 \times 10^{-29} ~ \Big(\frac{51}{d} \Big)^{2} ~ L_{x} ~ [\mu Jy],
\end{equation}
where F$_{3.5\, \rm cm}$ is the 3.5 cm flux, d is the distance to the source in pc and L$_{X}$ is the X-ray luminosity in $\rm erg~s^{-1}$, we find an expected flux of 4.3 $\mu$Jy. Free-free emission should have an approximately flat spectral index meaning the expected flux at 15 GHz (2 cm) should also be $\sim4.3~\mu$Jy. It is therefore possible that free-free emission accounts for up to $10\%$ of the observed 15 GHz flux. We note though that since the post-outburst X-ray luminosity should be lower than that observed by \citet{grosso10}, the expected post-outburst radio flux should be lower as well.

Considering all of the potential sources of the 15 GHz emission, we find that the most likely scenario is a combination of (gyro)synchrotron and free-free emission. We note though that follow-up observations at lower frequencies are needed to confirm the spectral index. It is indeed possible that there is a combination of more emission mechanisms present at 15 GHz. In order to disentangle all the potential sources of emission, $1-15$ GHz observations with both high angular resolution and sensitivity are necessary. The VLA is currently the leading facility in this regime and the observations presented are already pushing the limits of its capabilities.  Therefore, proposed future facilities, such as the ngVLA \citep{white18}, will be key to fully understanding the underlying emission mechanisms in EX Lupi and connecting them in the broader context of EXor/FUors in general.

\section{Summary}\label{sec:sum}

In this paper, we presented ALMA and VLA continuum observations of the EX Lupi disk in its post-outburst state. We fit radiative transfer models of the circumstellar dust and find the models are consistent with grain growth up to at least 3 mm, and possibly as high as 1 cm. The grain growth could have been spurred by the recent outburst, which ended in 2008. The best fit disk model has an ice line located at $0.2-0.3$ au in the disk's midplane, and the associated cooling timescales show there as been adequate time for the ice line to migrate inward from the observed position during the outburst. The most probable value for the total disk mass is 0.01 M$_{\odot}$, accompanied by a relatively compact characteristic dust radius of 45 au. The size is in agreement with other studies that find EXor/FUor disks to be more compact than that of typical protoplanetary disks. 

We find a best fit flux of 0.055 mJy at 15 GHz, significantly more than can be explained by thermal disk emission alone. We explored several potential sources of the emission and conclude that it is likely primarily due to (gyro)synchrotron emission coming from strong stellar magnetic fields and/or non-negligible free-free emission from accretion shocks and disk heating through X-ray emission.

\acknowledgments

We thank the anonymous referee for feedback that improved this paper. This project received support from the European Research Council (ERC) under the European Union's Horizon 2020 research and innovation program under grant agreement No 716155 (SACCRED). DS acknowledges support by the Deutsche Forschungsgemeinschaft through SPP 1833: ``Building a Habitable Earth'' (SE 1962/6-1). EV and VA acknowledge the support of the Large Scientific Project of the Russian Ministry of Science and Higher Education ``Theoretical and experimental studies of the formation and evolution of extrasolar planetary systems and characteristics of exoplanets'' (No. 13.1902.21.0039). On behalf of the SACCRED project we are thankful for the usage of MTA Cloud (https://cloud.mta.hu/) that helped us achieving the results published in this paper. This paper makes use of the ALMA data from projects ADS/JAO.ALMA\#2017.1.00224.S and ADS/JAO.ALMA\#2015.1.00200.S.  ALMA is a partnership of ESO (representing its member states), NSF (USA) and NINS (Japan), together with NRC (Canada), MOST and ASIAA (Taiwan), and KASI (Republic of Korea), in cooperation with the Republic of Chile. The Joint ALMA Observatory is operated by ESO, AUI/NRAO and NAOJ. The National Radio Astronomy Observatory is a facility of the National Science Foundation operated under cooperative agreement by Associated Universities, Inc. 


\vspace{5mm}
\facilities{ALMA, VLA}

\software{{\scriptsize CASA 5.4.1} \citep{casa_reference} 
          }

\end{document}